\documentclass[11pt,a4paper]{article}
\usepackage{}
\usepackage{bbm}
\usepackage{amssymb}
\usepackage{indentfirst,mathrsfs}
\usepackage{amsfonts,amsmath,amssymb,amsthm}
\usepackage{latexsym,amscd,multirow}
\usepackage{amsbsy}
\usepackage{color}
\usepackage[all,cmtip]{xy}
\usepackage{graphicx}
\newtheorem{thm}{Theorem}
\newtheorem{lem}{Lemma}
\newtheorem{cor}{Corollary}

\newtheorem{prop}{Proposition}

\newtheorem{example}{Example}
\newtheorem{remark}{Remark}

\newtheorem{conj}{Conjecture}

\newcounter{alg}
\newlength{\lefttab}
\newlength{\numberoffset}
{\trivlist
   \topsep=0pt\parsep=0pt\itemsep=0pt
   \addtolength{\lefttab}{1.25em}
   \addtolength{\numberoffset}{1.25em}
   \leftskip=\lefttab}%
  {\endtrivlist}

\begin{document}
\title{New Generalized Cyclotomic Binary Sequences of Period $p^2$}
\author{Zibi Xiao
\thanks{Z. Xiao and X. Zeng are with the Faculty of Mathematics and Statistics, Hubei Key Laboratory of Applied Mathematics,
Hubei University, Wuhan 430062, Hubei, China. Email: holly\_xzb@126.com, xzeng@hubu.edu.cn},
Xiangyong Zeng, Chunlei Li
\thanks{C. Li and T. Helleseth are with the Department of Informatics, University of Bergen,
Bergen {\rm N-5020}, Norway. Email: chunlei.li@uib.no, tor.helleseth@uib.no},
and
Tor Helleseth
}

\date{}
\maketitle
\begin{quote}
{\small {\bf Abstract:}} New generalized cyclotomic binary sequences of period $p^2$ are proposed in this paper, where $p$ is an odd prime.
The sequences are almost balanced and their linear complexity is determined. The result shows that the proposed
sequences have very large linear complexity if $p$ is a non-Wieferich prime.


\end{quote}
\section{Introduction}

The linear complexity of a sequence is defined to be the length of the shortest linear feedback shift register that can generate the sequence.
It is an important measure of randomness of sequences in stream ciphers. To resist the attack from the application of the Berlekamp-Massey algorithm,
the sequences used in cipher systems should have large linear complexity. Cyclotomic generators based on cyclotomy can generate sequences with larger linear complexity.

For an integer $n\geq 2$, let $\mathbb{Z}_n=\{0,1,\cdots,n-1\}$ denote the residue class ring of integers modulo $n$
and $\mathbb{Z}_n^*$ be the multiplicative group consisting of all invertible elements in $\mathbb{Z}_n$.
A partition $\{D_0^{(n)},D_1^{(n)}, \cdots, D_{d-1}^{(n)}\}$ of $\mathbb{Z}_n^*$ is a family of sets satisfying
$D_i^{(n)}\cap D_j^{(n)}=\emptyset$ for all $i\neq j$ and $\bigcup_{i=0} ^{d-1}D_i^{(n)} = \mathbb{Z}_n^*$. Suppose $D_0^{(n)}$ is a multiplicative subgroup
of $\mathbb{Z}_n ^*$ and there exist elements $g_i\in \mathbb{Z}_n ^*$ such that $D_i^{(n)}=g_iD_0^{(n)}$ for all $i=1,2,\cdots,d-1$,
the $D_i^{(n)}$ are called \emph{classical cyclotomic classes} of order $d$ with respect to $n$ when $n$ is prime,  and
\emph{generalized cyclotomic class} of order $d$ with respect to $n$ when $n$ is composite.

Legendre sequences, Ding-Helleseth-Lam sequences and Hall's sextic residue sequences are actually cyclotomic sequences based on classical
cyclotomy of order $2$, $4$ and $6$, respectively, and their linear complexities were also determined
\cite{Turyn, Ding3,Ding6, Kim1}.
Generalized cyclotomy with respect to $pq$ was presented by Whiteman for the purpose of searching for residue difference sets \cite{Whiteman},
and generalized cyclotomy with respect to $p^2$ was considered for cryptographic purpose \cite{Ding1}.
Whiteman's generalized cyclotomy of order $2$ was extended to the case with respect to $p_1^{e_1}\dots p_t^{e_t}$ \cite{Ding5}.
However, Whiteman's generalized cyclotomy is not consistent with classical cyclotomy.
A new generalized cyclotomy with respect to $p_1^{e_1}\dots p_t^{e_t}$ which includes classic cyclotomy as a special case was introduced in \cite{Ding4},
and it is referred to as Ding-Helleseth's generalized cyclotomy.
Thereafter Whiteman's and Ding-Helleseth's generalized cyclotomy were unified by an approach presented in \cite{Fan}.
To construct optimal frequency hopping sequences, another generalized cyclotomy with respect to $p_1^{e_1}\dots p_t^{e_t}$ was proposed in \cite{Zeng}.

The construction of generalized cyclotomic sequences is a natural application of {generalized cyclotomy.
Based on Whiteman's generalized cyclotomy of order 2, Ding in \cite{Ding2} presented a class of generalized cyclotomic sequences with period $pq$
and calculated their linear complexity. These sequences are not balanced. In \cite{Ding4}, Ding and Helleseth constructed a class of almost balanced binary
sequences via their own generalized cyclotomic classes of order $2$, and the linear complexity of the sequences with a specific period $pq$ was determined
in \cite{Bai}. Subsequently, the linear complexities of generalized cyclotomic sequences with different period have been extensively studied in the literature
\cite{Kim2}, \cite{Yan1}, \cite{Yan2}, \cite{Hu} and \cite{Ke}. Recently, a generalization of Hall's sextic residue sequences was proposed and
the linear complexity was also determined \cite{Du}.

In this paper, new binary sequences of period $p^2$ for an odd prime $p$ are constructed based on the generalized cyclotomy introduced in \cite{Zeng}.
Different from the previous ones in \cite{Ding4, Ding5, Fan}, the order of the generalized cyclotomy in \cite{Zeng} depends on the choice of parameters.
The flexibility in choosing the parameters allows us to construct a large number of cyclotomic sequences for a fixed primitive element modulo $p^2$.	
In order to determine the linear complexity of the constructed cyclotomic sequences $(s_0,s_1,\cdots, s_{p^2-1})$,
we adopt the classic approach described in \cite{Ding2, Ding3} and focus on the investigation of the roots of the polynomial
$\sum_{i=0}^{p^2-1}s_ix^i$ in the splitting field of $x^{p^2}-1$ over the finite field $\mathbb{F}_2$.
Based on careful manipulations of properties of the generalized cyclotomy in \cite{Zeng},
we derive a system of equations from some of the cyclotomic classes. The analysis of the number of solutions to the system enables us to characterize necessary and sufficient
conditions for the polynomial $\sum_{i=0}^{p^2-1}s_ix^i$ to share roots with $x^{p^2}-1$, and to eventually determine  the linear complexity of the new generalized cyclotomic sequences.
The linear complexity of the constructed cyclotomic sequences is, in general, larger than that of the generalized cyclotomic
sequences of period $p^2$ in \cite{Ding0}, \cite{Yan1} and \cite{Du}.
For small primes $p$, the numerical result indicates that
the cyclotomic sequences are  ``good"  in terms of
the linear complexity profile \cite{Niederreiter}, the well-distribution measure and the correlation measure \cite{Mauduit}
of pseudorandom sequences.

The remainder of this paper is organized as follows. Section $2$ proposes a construction of generalized cyclotomic sequences. The linear complexity of
the constructed sequences is determined in Section $3$. Section $4$ concludes the study.

\section{Preliminaries}

In this section, we first recall the generalized cyclotomic classes with respect to $p^j$ for any integer $j\geq 1$,
and then define the new generalized cyclotomic sequences of period $p^2$.

Let $p$ be an odd prime and $g$ be a primitive root modulo $p^2$. It is well known that $g$ is also a
primitive root modulo $p^j$ for each integer $j\geq 1$ \cite{Apostol}. Denote $p=ef+1$.
For an integer $j\geq 1$, let $d_j=\varphi(p^j)/e=p^{j-1}f$, where $\varphi(\cdot)$ is the Euler's totient function. Define
\begin{equation}\label{Eq_GenCyclotomy}
\begin{array}{l}
D_0^{(p^j)} = \{g^{d_jt}(\textrm{mod}\, p^j)\,:\,0\leq t< e\},  \text{ and} \\
D_i^{(p^j)} = g^iD_0^{(p^j)}=\{g^ix(\textrm{mod}\, p^j)\,:\,x\in D_0^{(p^j)}\},\,\,i=1, 2,\cdots, d_j-1.
\end{array}
\end{equation}
It was shown in \cite{Zeng} that $\{D_0^{(p^j)}, D_1^{(p^j)}, \cdots, D_{d_j-1}^{(p^j)}\}$ forms a partition of $\mathbb{Z}_{p^j}^*$
and $D_i^{(p^j)}$ with $0\leq i\leq d_j-1$} are generalized cyclotomic classes of order $d_j$ with
respect to $p^j$.
Moreover, for an integer $m\geq 1$,
\begin{equation*}
\mathbb{Z}_{p^m}\backslash \{0\} = \bigcup\limits_{j=1}^m p^{m-j}\mathbb{Z}_{p^j}^*=\bigcup\limits_{j=1}^m \bigcup\limits_{i=0}^{d_j-1} p^{m-j}D_i^{(p^j)}.
\end{equation*}

In the following we shall define a family of generalized cyclotomic sequences of period $p^2$.

Take $f=2^r$ with $r\geq 1$ and let $b$ be an integer with $0\leq b\leq pf-1$.
Note that $$\mathbb{Z}_{p^2}\backslash \{0\}=p\mathbb{Z}_{p}^*\cup \mathbb{Z}_{p^2}^*=\bigcup\limits_{i=0}^{f-1}pD_i^{(p)}\bigcup\bigcup\limits_{i=0}^{pf-1} D_i^{(p^2)}.$$
Define two sets \[C_0=\bigcup\limits_{i=f/2}^{f-1}pD_{(i+b)\,(\textrm{mod}\,f)}^{(p)}\bigcup\limits_{i=pf/2}^{pf-1} D_{(i+b)\,(\textrm{mod}\,pf)}^{(p^2)},\]
\begin{equation}\label{e0}
C_1=\bigcup\limits_{i=0}^{f/2-1}pD_{(i+b)\,(\textrm{mod}\,f)}^{(p)}\bigcup\limits_{i=0}^{pf/2-1} D_{(i+b)\,(\textrm{mod}\,pf)}^{(p^2)}\cup \{0\}.
\end{equation}
It is straightforward that $\{C_0, C_1\}$ is a partition of $\mathbb{Z}_{p^2}$ and $|C_1|-|C_0|=1$.
We now define a family of generalized cyclotomic binary sequences of period $p^2$ that admits
$C_1$ as the \textit{characteristic set}, i.e.,
the sequences $\mathbf{s}^\infty=(s_0,s_1,s_2,\cdots )$  are given by
\begin{equation}\label{e1}
s_i = \left\{ {\begin{array}{*{20}c}
   {0}, \hfill & {i\,(\textrm{mod}\,p^2) \in C_0,} \hfill  \\
   {1}, \hfill & {i\,(\textrm{mod}\,p^2) \in C_1.} \hfill  \\
\end{array}} \right.
\end{equation}

\section{Linear complexity of the sequence  $\mathbf{s}^\infty$}
Let $\mathbf{s}^\infty=(s_0,s_1,s_2,\cdots )$ be a binary sequence of period $n$ and
$s(x)=s_0+s_1x+\cdots+s_{n-1}x^{n-1}$.
It is well known (see, for example, \cite[page 171]{Cusick}) that the \textit{minimal polynomial} of
$\mathbf{s}^\infty$ is  $(x^n-1)/\gcd(x^n-1, s(x))$ and
the \textit{linear complexity} of  $\mathbf{s}^\infty$ is given by
\begin{equation}\label{e2}
L(\mathbf{s}^{\infty})= n-\textrm{deg}(\textrm{gcd}(x^n-1,s(x))).
\end{equation}
In this section, we shall determine the linear complexity of the new generalized cyclotomic binary sequences of period $p^2$ defined in (\ref{e1}).
Let $\beta$ be an arbitrary primitive $p^2\textrm{th}$ root of unity in the splitting field of $x^{p^2}-1$ over $\mathbb{F}_2$. Then (\ref{e2}) can be rewritten as
\begin{equation}\label{e4-0}
L(\mathbf{s}^{\infty})= p^2-|\,\{a \in \mathbb{Z}_{p^2}\,:\,s(\beta^a) = 0\}\,|.
\end{equation}
Hence the problem of determining the linear complexity of the sequence defined by (\ref{e1}) is translated to that of investigating the cardinality of the set
$
\{a \in \mathbb{Z}_{p^2}\,:\,s(\beta^a) = 0\}.
$

For brevity, we define
\begin{equation}\label{e5}
H_{b\,(\textrm{mod}\,f)}^{(p)}=\bigcup\limits_{i=0}^{f/2-1}pD_{(i+b)\,(\textrm{mod}\,f)}^{(p)}, \,\,
H_b^{(p^2)}=\bigcup\limits_{i=0}^{pf/2-1} D_{(i+b)\,(\textrm{mod}\,pf)}^{(p^2)}.
\end{equation}
In the rest of this paper, the subscripts $i$ in $D_i^{(p^{j})}$ and $H_i^{(p^{j})}$ will be always assumed to be taken modulo the order
$d_j=p^{j-1}f$, and the modulo operation will be omitted for simplicity.
Thus the characteristic set $C_1$ of the sequence defined by (\ref{e1}) can be written as
$$C_1= H_{b}^{(p)}\cup H_b^{(p^2)} \cup \{0\}.$$
Throughout what follows,  we denote by $T(x)$ the polynomial $\sum_{t\in T}x^t$ for any set $T$.
Then for the sequence $\mathbf{s}^{\infty}$ defined in (\ref{e1}), we have
\begin{equation}\label{e3}
      s(x)=\sum\limits_{t=0}^{p^2-1}{s_ix^t} = C_1(x)=1+H_b^{(p)}(x)+H_b^{(p^2)}(x).
\end{equation}
Thus, $$ L(\mathbf{s}^{\infty})=p^2 - \Big|\,\big\{a \in \mathbb{Z}_{p^2}\,:\,1+H_b^{(p)}(\beta^a)+H_b^{(p^2)}(\beta^a) = 0\big\}\Big|.$$

To calculate the linear complexity of $\mathbf{s}^{\infty}$ defined in \eqref{e1}, we begin with presenting some properties of the generalized cyclotomy given in \eqref{Eq_GenCyclotomy} and useful lemmas in Subsection 3.1.

\subsection{Useful lemmas}


We first characterize properties of the generalized cyclotomy with respect to $p^j$ for $j\in\{1,2\}$.
\begin{lem}\label{lem2} With the notation in (\ref{Eq_GenCyclotomy}), we have:

(i) $aD_i^{(p)}(\bmod\,p)=D_{i+k}^{(p)}$ for $a\in D_k^{(p)}$, and  $aD_i^{(p^2)}(\bmod\,p^2)=D_{i+k}^{(p^2)}$ for $a\in D_k^{(p^2)}$;

(ii) if $a\in D_k^{(p^2)}$, then $aD_i^{(p)}(\bmod\,p)=D_{i+k}^{(p)}$;

(iii) if $a\in pD_{k}^{(p)}$, then $apD_i^{(p)}(\bmod\,p^2)=\{0\}$, $aD_i^{(p^2)}(\bmod\,p^2)=pD_{i+k}^{(p)}$.
\end{lem}
\begin{proof}
The assertion in (i) can be easily verified, and so we only prove (ii) and (iii) here. Since $a\in D_k^{(p^2)}$, there exists
a uniquely determined integer $t_0$ with $0\leq t_0\leq e-1$ such that $a \equiv g^{pft_0+k}(\bmod\,p^2)$.
It is clear that $a \equiv g^{pft_0+k}(\bmod\,p)$. Due to $g^{p-1}\equiv 1(\bmod\,p)$, we have
$a\equiv g^{(p-1)ft_0+ft_0+k}\equiv g^{ft_0+k}(\bmod\,p)$ and then
\begin{eqnarray*}
  aD_i^{(p)}(\bmod\,p) &=& \{(g^{ft_0+k}\cdot g^{ft+i})(\bmod\,p)\,\,:\,\,0\leq t\leq e-1\}\\
              &=& \{g^{f(t_0+t)+i+k}(\bmod\,p)\,\,:\,\,0\leq t\leq e-1\}\\
              &=& D_{i+k}^{(p)}.
\end{eqnarray*}
For the assertion in (iii),
since $a\in pD_{k}^{(p)}$, $a$ can be expressed as $a=pg^{ft_0+k}(\textrm{mod}\,p^2)$. Any $u \in pD_i^{(p)}$ can be expressed
as $u=pg^{ft+i}(\textrm{mod}\,p^2)$. As a consequence, $au=p^2g^{f(t+t_0)+i+k} \equiv 0\,(\textrm{mod}\,p^2)$ and then $apD_i^{(p)}(\textrm{mod}\,p^2)=\{0\}$.
Since $g^{p-1}=g^{fe}\equiv 1\,(\textrm{mod}\,p)$, we have
\begin{eqnarray*}
  aD_i^{(p^2)}(\textrm{mod}\,p^2)  &=& \{(pg^{ft_0+k}\cdot g^{pft+i})(\textrm{mod}\,p^2)\,{\color{red}\,:\,}\,0\leq t\leq e-1\}\\
              &=& p\cdot \{g^{(p-1)ft+ft+ft_0+i+k}(\textrm{mod}\,p)\,{\color{red}\,:\,}\,0\leq t\leq e-1\}\\
              &=& p\cdot\{g^{f(t+t_0)+i+k}(\textrm{mod}\,p)\,{\color{red}\,:\,}\,0\leq t\leq e-1\}\\
              &=& pD_{i+k}^{(p)}.
\end{eqnarray*}
\end{proof}

Next we derive the formula for $H_b^{(p^j)}(\beta^a)$ with $j\in \{1,2\}$. It is necessary to consider the cases $a\in D_k^{(p^2)}$ and $a\in pD_k^{(p)}$ separately.
\begin{lem}\label{lem3}
If $a\in D_k^{(p^2)}$, then $H_b^{(p^j)}(\beta^a)=H_{b+k}^{(p^j)}(\beta)$ for $j = 1, 2.$
\end{lem}
\begin{proof}
Since $a\in D_k^{(p^2)}$, we have $aD_i^{(p)}(\textrm{mod}\,p)=D_{i+k}^{(p)}$ and $aD_i^{(p^2)}(\textrm{mod}\,p^2)=D_{i+k}^{(p^2)}$
by Lemma \ref{lem2} (ii) and (i), respectively. Moreover, by the definitions of sets $H_b^{(p^2)}$ and $H_{b}^{(p)}$,
\[
aH_b^{(p^2)}(\textrm{mod}\,p^2)=\bigcup\limits_{i=0}^{pf/2-1} aD_{i+b}^{(p^2)}(\textrm{mod}\,p^2)=\bigcup\limits_{i=0}^{pf/2-1}D_{i+b+k}^{(p^2)}=H_{b+k}^{(p^2)},\]
\[aH_{b}^{(p)}(\textrm{mod}\,p^2)=p\bigcup\limits_{i=0}^{f/2-1} aD_{i+b}^{(p)}(\textrm{mod}\,p)=\bigcup\limits_{i=0}^{f/2-1} pD_{i+b+k}^{(p)}=H_{b+k}^{(p)}.\]
Therefore, for $j=1,2$, we have
\[H_{b}^{(p^j)}(\beta^a)=\sum\limits_{t\in H_{b}^{(p^j)}}{(\beta^a)^t}=\sum\limits_{t\in H_{b+k}^{(p^j)}}{\beta^t}=H_{b+k}^{(p^j)}(\beta).\]
\end{proof}

\begin{lem}\label{lem5}
For $a\in pD_k^{(p)}$,
$
H_b^{(p)}({\beta^a})=\frac{p-1}{2}\,(\bmod\,2)$ and $H_b^{(p^2)}({\beta^a})=\frac{{p-1}}{2}\,(\bmod{\,2})+H_{b+k}^{(p)}(\beta).
$
\end{lem}

\begin{proof}
Since $a\in pD_k^{(p)}$, for any $u\in pD_{i}^{(p)}$, the congruence $au\equiv 0\,(\bmod\,p^2)$ holds by Lemma \ref{lem2} (iii).
It can be verified that \[H_{b}^{(p)}(\beta^a)=\sum\limits_{t\in H_{b}^{(p)}}{(\beta^a)^t}=\frac{{p-1}}{2}(\bmod{\,2}).\]
In addition, we have $aD_i^{(p^2)}(\bmod\,p^2)=pD_{i+k}^{(p)}$ by Lemma \ref{lem2} (iii) and then
\[aH_b^{(p^2)}(\bmod\,p^2)=\bigcup\limits_{i=0}^{pf/2-1} aD_{i+b}^{(p^2)}(\bmod\,p^2)=\bigcup\limits_{i=0}^{pf/2-1} pD_{i+b+k}^{(p)}.\]
When $i$ runs through $\{\,0,1,\,\cdots,\,pf/2-1\}$, $i\,(\textrm{mod}\,f)$ takes on each element of
$\{\,0,1,\,\cdots,\,f-1\}$ $(p-1)/2$ times and that of $\{\,b,b+1,\,\cdots,\,b+f/2-1\}$ one additional time.
Therefore,
\begin{eqnarray*}
  H_b^{(p^2)}(\beta^a) &=& \sum\limits_{t\in H_b^{(p^2)}}{(\beta^a)^t}\\
                  &=& \frac{{p-1}}{2}\sum\limits_{t\in p\mathbb{Z}_p^*}{\beta^t}+\sum\limits_{t\in H_{b+k}^{(p)}}{\beta^t}\\
                                 &=& \frac{{p-1}}{2}\,(\textrm{mod}\,2)+H_{b+k}^{(p)}(\beta).
\end{eqnarray*}
\end{proof}

Applying Lemmas \ref{lem3} and \ref{lem5}, we then obtain an expression for $s({\beta^a})$.
\begin{prop}\label{lem6}
Let $D_k^{(p^j)}$ and $H_{b+k}^{(p^j)}$ for $j=1, 2$ be defined as in (\ref{Eq_GenCyclotomy}) and (\ref{e5}), respectively. Then for $0\leq k\leq pf-1$, we have
\[
s(\beta^a) = \left\{ {\begin{array}{*{20}c}
   1, \hfill & a = 0, \hfill  \\
   1+ H_{b+k}^{(p)}(\beta), \hfill & {a \in pD_{k}^{(p)}}, \hfill  \\
    1+ H_{b+k}^{(p)}(\beta)+H_{b+k}^{(p^2)}({\beta}), \hfill & {a \in D_k^{(p^2)}}. \hfill  \\
\end{array}} \right.
\]
\end{prop}

\begin{proof}
The results for the case $a\in D_k^{(p^2)}$ and $a \in pD_k^{(p)}$ follow directly from (\ref{e3}), Lemmas \ref{lem3} and \ref{lem5}. For the case $a=0$, we have $s(\beta^a)=|C_1|=\frac{p^2+1}{2}(\bmod\,2)=1$.
\end{proof}

By Proposition \ref{lem6}, we need to further investigate the value of $H_{b+k}^{(p^j)}(\beta)$
for $j = 1, 2$.
 The following property of the primitive $p^2$th root of unity $\beta$ is of great use in the sequel.

\begin{lem}\label{lem1}For any integer $b$ with $0\leq b\leq pf-1$, we have
\[H_b^{(p)}(\beta)+H_{b+f/2}^{(p)}(\beta)=\sum\limits_{t\in p\mathbb{Z}_p^*}\beta^t=1,\,\,\,\,H_b^{(p^2)}(\beta)+H_{b+pf/2}^{(p^2)}(\beta)=\sum\limits_{t\in \mathbb{Z}_{p^2}^*}\beta^t=0.\]
\end{lem}

\begin{proof}
Note that $\beta$ is a primitive $p^2\textrm{th}$ root of unity in an extension field of $\mathbb{F}_2$ and \[0=\beta^{p^2}-1=(\beta-1)(\beta^{p^2-1}+\beta^{p^2-2}+\cdots+\beta+1),\] \[0=\beta^{p^2}-1=(\beta^p-1)(\beta^{p(p-1)}+\beta^{p(p-2)}+\cdots+\beta^p+1).\]
Therefore, \[\sum\limits_{t=0}^{p-1}{(\beta^p)^t}=0, \,\,\,\,\,\,\sum\limits_{t=0}^{p^2-1}{\beta^t}=0.\]
By the definitions of sets $H_b^{(p)}$ and $H_{b}^{(p^2)}$, we have
\[H_b^{(p)}\cup H_{b+f/2}^{(p)}=\bigcup\limits_{i=0}^{f-1} pD_i^{(p)}=pZ_p^*, \,\,\,\,\,\,
H_{b}^{(p^2)}\cup H_{b+pf/2}^{(p^2)}=\bigcup\limits_{i=0}^{pf-1} D_i^{(p^2)}=Z_{p^2}^*.\]
Consequently,
\[H_b^{(p)}(\beta)+H_{b+f/2}^{(p)}(\beta)=\sum\limits_{t\in pZ_p^*}{\beta^t}=\sum\limits_{t\in Z_p^*}{(\beta^p)^t}=\sum\limits_{t=1}^{p-1}{(\beta^p)^t}=1,\]
and \[H_{b}^{(p^2)}(\beta)+H_{b+pf/2}^{(p^2)}(\beta)=\sum\limits_{t\in Z_{p^2}^*}{\beta^t}=\sum\limits_{t=0}^{p^2-1}{\beta^t}-\sum\limits_{t=0}^{p-1}{(\beta^p)^t}=0.\]
\end{proof}

We are now ready to investigate the conditions under which the value of $s(\beta^a)$ might be zero. By Proposition \ref{lem6}, it suffices to consider the values of $H_v^{(p)}(\beta)$ for each $v$ with $0\leq v\leq f-1$ and $H_v^{(p)}(\beta)+H_v^{(p^2)}(\beta)$ for each $v$ with $0\leq v\leq pf-1$.

\begin{lem}\label{lem7}For any integer $v$ with $0\leq v\leq f-1$,
$H_v^{(p)}(\beta) \in \{0,1\} \text{ if and only if }2 \in D_0^{(p)}.$
\end{lem}
\begin{proof}
For the sufficiency of the claim, if $2 \in D_0^{(p)}$, then $2D_i^{(p)}(\bmod\,p)=D_{i}^{(p)}$ by Lemma \ref{lem2}. It follows from the definition of $H_v^{(p)}$ that
\[2H_v^{(p)}(\bmod\,p)=p\bigcup\limits_{i=0}^{f/2-1} 2D_{i+v}^{(p)}\,(\bmod\,p)=\bigcup\limits_{i=0}^{f/2-1} pD_{i+v}^{(p)}=H_v^{(p)}.\]
Hence \[(H_v^{(p)}(\beta))^2=H_v^{(p)}(\beta^2)=\sum\limits_{t\in H_v^{(p)}}{(\beta^2)^t}=\sum\limits_{t\in H_v^{(p)}}{\beta^t}=H_v^{(p)}(\beta),\]
and then $H_v^{(p)}(\beta) \in \{0,1\}$ for each $v$, $0\leq v\leq f-1$.

For the necessity, suppose that
 $2\notin D_0^{(p)}$, say $2 \in D_u^{(p)}$ with $0<u\leq f-1$. Since
 $H_v^{(p)}(\beta) \in \{0,1\}$, we have
\[H_v^{(p)}(\beta) = (H_v^{(p)}(\beta))^{2^n}= H_v^{(p)}(\beta^{2^n})= H_{(v+nu)(\textrm{mod}\,f)}^{(p)}(\beta)\]
for any positive integer $n$.
 We note that $f=2^r$ implies gcd$(u,f)=2^i$ with $0 \leq i\leq r-1$, and so gcd$(u,f)|\frac{f}{2}$.
Hence, the congruence $ux\equiv f/2\,(\textrm{mod}\,f)$ has solutions, and thus
$H_v^{(p)}(\beta)=H_{v+f/2}^{(p)}(\beta)$. However, by Lemma \ref{lem1}, we have
$H_v^{(p)}(\beta)+H_{v+f/2}^{(p)}(\beta)=1$ for all $v$, $0\leq v\leq f-1$. This leads to a contradiction and the proof is completed.
\end{proof}

\begin{lem}\label{lem8}
Let $A_v(\beta)=H_v^{(p)}(\beta)+H_v^{(p^2)}(\beta)$. Then for any integer $v$ with $0\leq v\leq pf-1$,  $A_v(\beta)\in \{0,1\}$  if and only if $2 \in D_0^{(p^2)}$.
\end{lem}
\begin{proof}
The sufficiency can be proved in a similar manner to Lemma \ref{lem7}, so we skip it and only prove the necessity here.

Given an integer $v$, $0\leq v\leq pf-1$, suppose $A_v(\beta)\in \{0,1\}$ and
$2\in D_u^{(p^2)}$ with $0<u\leq pf-1$. From the fact $\frac{pf}{2}\equiv \frac{f}{2}(\textrm{mod}\,f)$ and Lemma \ref{lem1} it follows that
\begin{equation}\label{e9}
    A_v(\beta)+A_{(v+\frac{pf}{2})(\textrm{mod}\,pf)}(\beta)=H_v^{(p)}(\beta)+H_v^{(p^2)}(\beta)+H_{v+\frac{pf}{2}}^{(p)}(\beta)+H_{v+\frac{pf}{2}}^{(p^2)}(\beta)=1,
\end{equation}
which implies $A_{(v+\frac{pf}{2})(\textrm{mod}\,pf)}(\beta)\in \{0,1\}$.
	Since $2\in D_u^{(p^2)}$ with $0<u\leq pf-1$,
	it follows from Lemma \ref{lem2}(i) that $2D_v^{(p^j)}=D_{v+u}^{(p^j)}$ for $j=1, 2$.
	We obtain in the same way as in the proof of Lemma \ref{lem7} that
\begin{equation}\label{e6}
A_v(\beta)=A_{(v+u)(\textrm{mod}\,pf)}(\beta)=\cdots=A_{(v+(p-1)u)(\textrm{mod}\,pf)}(\beta){\color{blue},}
\end{equation}
and
\begin{equation}\label{e7}
A_{(v+\frac{pf}{2})(\textrm{mod}\,pf)}(\beta)=A_{(v+\frac{pf}{2}+u)(\textrm{mod}\,pf)}(\beta)=\dots=A_{(v+\frac{pf}{2}+(p-1)u)(\textrm{mod}\,pf)}(\beta).
\end{equation}
We distinguish now two cases.

\emph{Case 1:} $u$ is not divisible by $f$.
Recall that $p$ is an odd prime and $f=2^r$ with $r\geq 1$. Hence gcd$(u,pf)$ is of the form $2^ip^j$ with
$0 \leq i\leq r-1, 0\leq j\leq 1$ and therefore divides $\frac{pf}{2}$. It then follows that
there exists an integer $n$, $0<n<p$, satisfying the congruence
$nu \equiv \frac{pf}{2}(\textrm{mod}\,pf)$. Together with (\ref{e6})
we get $A_v(\beta)=A_{(v+\frac{pf}{2})(\textrm{mod}\,pf)}(\beta)$, a contradiction to (\ref{e9}).

\emph{Case 2:} $u$ is divisible by $f$, say $u=fd$ with $0\leq d\leq p-1$. Let $v\equiv h\,(\textrm{mod}\,f)$ with $0\leq h\leq f-1$.
Writing $v=kf+h$ with $k\in \mathbb{N}$ and noting that gcd$(d,p)=1$, we get \[\{(v+nu)(\textrm{mod}\,pf)\,{\color{red}\,:\,}\,0\leq n\leq p-1\}
=\{h+nf\,{\color{red}\,:\,}\,0\leq n\leq p-1\}.\]
Therefore, (\ref{e6}) can be written as
\[A_h(\beta)=A_{h+f}(\beta)=\dots=A_v(\beta)=\dots=A_{h+(p-1)f}(\beta).\]
Similarly, (\ref{e7}) can be written as
\[A_{h+\frac{f}{2}}(\beta)=A_{h+\frac{3f}{2}}(\beta)=\dots=A_{(v+\frac{pf}{2})(\textrm{mod}\,pf)}(\beta)=\dots=A_{h+\frac{f}{2}+(p-1)f}(\beta).\]

By combining the above identities, it follows that for $n=0,1,\cdots,p-1$,
\[A_{h+\frac{nf}{2}}(\beta)+A_{h+\frac{nf}{2}+\frac{f}{2}}(\beta)=A_v(\beta)+A_{(v+\frac{pf}{2})(\textrm{mod}\,pf)}(\beta)=1,\]
which together with $H_{h+\frac{nf}{2}}^{(p)}(\beta)+H_{h+\frac{nf}{2}+\frac{f}{2}}^{(p)}(\beta)=1$ implies
$H_{h+\frac{nf}{2}}^{(p^2)}(\beta)+H_{h+\frac{nf}{2}+\frac{f}{2}}^{(p^2)}(\beta) = 0$. From the
definition of $H_b^{(p^2)}$, one has
\[
H_{h+\frac{nf}{2}}^{(p^2)}\cap H_{h+\frac{nf}{2}+\frac{f}{2}}^{(p^2)}
=\bigcup\limits_{i=\frac{f}{2}}^{\frac{pf}{2}-1}D_{i+h+\frac{nf}{2}}^{(p^2)}.
\] Thus,
\[
H_{h+\frac{nf}{2}}^{(p^2)}(\beta)+H_{h+\frac{nf}{2}+\frac{f}{2}}^{(p^2)}(\beta) = \sum\limits_{i=0}^{\frac{f}{2}-1}D_{i+h+\frac{nf}{2}}^{(p^2)}(\beta) + \sum\limits_{i=0}^{\frac{f}{2}-1}D_{i+h+\frac{pf}{2}+\frac{nf}{2}}^{(p^2)}(\beta) = 0.
\]
For $0\leq n\leq p-1$, let $$U_n=\bigcup\limits_{i=0}^{\frac{f}{2}-1}\Big(D_{i+h+\frac{nf}{2}}^{(p^2)}\cup D_{i+h+\frac{pf}{2}+\frac{nf}{2}}^{(p^2)}\Big).$$ Then
the polynomial $U_n(x) = \sum_{t\in U_n} x^t$ in $\mathbb{F}_2[x]$ vanishes on the primitive $p^2$th root of unity $\beta$.


Note that
each polynomial $U_n(x)$ with $0\leq n\leq p-1$
satisfies $U_n(\beta)=0$ for an arbitrary primitive $p^2$th root of unity $\beta$.
This implies the degree of $U_n(x)$ is no less than the number of primitive $p^2$th roots of unity
over $\mathbb{F}_2$. Furthermore, the fact $U_n\subseteq \mathbb{Z}^*_{p^2}$ implies
$\deg(U_n(x))> p(p-1)$ for each $n$ with $0\leq n \leq p-1.$

It is easy to check that
 $\{U_n\,|\,0 \leq n \leq p-1\}$ forms a partition of $\mathbb{Z}_{p^2}^*$. On one hand,
 the sets $U_n$, $0\leq n \leq p-1,$
 yield $p$ polynomials $U_0(x), U_1(x), \cdots, U_{p-1}(x)$,
 among which the terms are pairwise distinct and each polynomial has degree larger $p(p-1)$.
 On the other hand,
there are in total only $(p-1)$ integers greater than $p(p-1)$ in $\mathbb{Z}_{p^2}^*$. This leads to a contradiction.
\end{proof}

We conclude this subsection with a necessary and sufficient condition for $2$ to be in the cyclotomic class $D_0^{(p^j)}$ for $j\in\{1,2\}$.

\begin{lem}\label{lem9} Let $D_0^{(p)}$ and $D_0^{(p^2)}$ be defined as in (\ref{Eq_GenCyclotomy}).
 Then  $2 \in D_0^{(p^j)}$ if and only if $2^e \equiv 1\,(\bmod\,p^j)$ for $j=1, 2$.
\end{lem}
\begin{proof}
If $2 \in D_0^{(p^j)}$, $j=1,2$, then there exists an integer $t_0$, $1\leq t_0 \leq e-1$, such that $2\equiv g^{p^{j-1}ft_0}(\textrm{mod}\,p^j)$.
Note that $g$ is a primitive root modulo $p^j$ and $ef=p-1$. Consequently, $2^e\equiv g^{p^{j-1}eft_0}\equiv g^{p^{j-1}(p-1)t_0}\equiv 1\,(\textrm{mod}\,p^j)$.
Conversely, suppose that $2^e\equiv 1\,(\textrm{mod}\, p^j)$ and $2\equiv g^{p^{j-1}ft_0+i}(\textrm{mod}\,p^j)$ for some $0\leq t_0 \leq e-1$ and $0\leq i\leq p^{j-1}f-1$.
Then $2^e\equiv g^{p^{j-1}eft_0+ei}\equiv g^{ei}\equiv 1\,(\textrm{mod}\,p^j)$, which implies $ei\equiv 0\,(\textrm{mod}\,\, p^{j-1}(p-1))$, and thus, $p^{j-1}f$ divides $i$. Since $0\leq i\leq p^{j-1}f-1$, we have $i=0$ and then $2 \in D_0^{(p^j)}$.
\end{proof}

\subsection{Main results}
After the preparations in Subsection 3.1, we can now summarize the main results of this paper in the following two theorems.
\begin{thm}\label{thm1}
Let $p$ be an odd prime and let $e$ be a factor of $p-1$ such that $f=\frac{p-1}{e}$ is of the form $2^r$ with $r\geq 1$.
Let $\mathbf{s}^\infty$ be a generalized cyclotomic binary sequence of period $p^2$ defined by (\ref{e1}).
If $2^e\not\equiv 1\,(\bmod\, p^2)$, then
\[L(\mathbf{s}^\infty) = \left\{ {\begin{array}{*{20}c}
   p^2-\frac{{p-1}}{2}, \hfill & {2 \in D_0^{(p)}}, \hfill  \\
   p^2, \hfill & {2 \notin D_0^{(p)}}. \hfill  \\
\end{array}} \right.\]
\end{thm}
\begin{proof}
For $a\in D_k^{(p^2)}$ with $0\leq k\leq pf-1$,  by Proposition \ref{lem6} we have
\[s(\beta^a)=1+H_{b+k}^{(p)}(\beta)+H_{b+k}^{(p^2)}({\beta}).\]
Since $2^e\not\equiv 1\,(\bmod\, p^2)$ implies $2\notin D_0^{(p^2)}$, it then follows
from Lemma \ref{lem8} that $s(\beta^a)\notin \{0,1\}$ for any $a\in \mathbb{Z}_{p^2}^*$.

For $a \in pD_{k}^{(p)}$ with $0\leq k\leq f-1$, $s(\beta^a)= 1+H_{b+k}^{(p)}(\beta)$ by Proposition \ref{lem6}.
We distinguish now the cases $2\in D_0^{(p)}$ and $2\notin D_0^{(p)}$. If $2\in D_0^{(p)}$, then by Lemma
\ref{lem7}, we have $H_{b+k}^{(p)}(\beta)\in \{0,1\}$ and hence $s({\beta^a})\in \{0,1\}$ for all
$a\in pD_{k}^{(p)}$. Since $H_{b+k}^{(p)}(\beta)+H_{b+k+\frac{f}{2}}^{(p)}(\beta)=1$ by Lemma \ref{lem1},
we have $s({\beta^a})=1$ for half of the elements in $ p\mathbb{Z}_p^*$ and $s({\beta^a})=0$ for the other half,
that is, there are exactly $\frac{p-1}{2}$ elements $a \in p\mathbb{Z}_p^*$ such that $s({\beta^a})=0$.
If $2\notin D_0^{(p)}$, then by Lemma \ref{lem7}, we have $H_{k}^{(p)}(\beta)\notin \{0,1\}$ for any
$0\leq k\leq f-1$, and thus $s({\beta^a})\notin \{0,1\}$ for any $a \in p\mathbb{Z}_p^*$.

Finally, we have $s({\beta^a})=1$ for $a=0$ again by Proposition \ref{lem6}. In conclusion, the size of the
set $\{a\in\mathbb{Z}_{p^2}\,:\,s(\beta^a)=0\}$ is $0$ if $2\notin D_0^{(p)}$ and $\frac{p-1}{2}$ if $2\in D_0^{(p)}$,
and so the result follows immediately from (\ref{e4-0}).
\end{proof}

If $f$ is taken to be $2$,
it was shown in \cite{Ding3} that $2\in D_0^{(p)}$ for an odd prime $p$
if and only if $p\equiv \pm1\,(\bmod\,8)$. So we obtain the following corollary.
\begin{cor}\label{cor1}
Let $p$ be an odd prime and $e=\frac{p-1}{2}$. If $2^e\not\equiv 1\,(\bmod\,p^2)$, then the linear complexity
of the generalized cyclotomic binary sequences $\mathbf{s}^{\infty}$ defined by (\ref{e1}) satisfies
\[L(\mathbf{s}^{\infty}) = \left\{ {\begin{array}{*{20}c}
   p^2-\frac{{p-1}}{2}, \hfill & {p\equiv \pm1\,(\bmod\,8)}, \hfill  \\
   p^2, \hfill & {p\equiv \pm3\,(\bmod\,8)}. \hfill  \\
\end{array}} \right.\]
\end{cor}

For the case $2^e\equiv 1\,(\bmod\, p^2)$, we have  the following result.
\begin{thm}\label{thm2}
Let $p$ be an odd prime and $e$ be a factor of $p-1$ such that $f=\frac{p-1}{e}$ is of the form $2^r$
with $r\geq 1$. If $2^e\equiv 1\,(\bmod\, p^2)$, then the linear complexity of the generalized cyclotomic binary sequences
defined by (\ref{e1}) is equal to $\frac{p^2+1}{2}$.
\end{thm}
\begin{proof}
Since $2^e\equiv 1\,(\bmod\,p^2)$ implies $2^e\equiv 1\,(\textrm{mod}\,p)$, it then follows from Lemma \ref{lem9} that $2 \in D_0^{(p^2)}$
and $2 \in D_0^{(p)}$. Therefore, by Proposition \ref{lem6}, Lemmas \ref{lem7} and \ref{lem8} we have
\[s(\beta^a)=1+H_{b+k}^{(p)}(\beta)\in \{0,1\}\,\,\,\, \textrm{for}\,\,\,\, a\in pD_k^{(p)},\]
and \[s(\beta^a)=1+H_{b+k}^{(p)}(\beta)+H_{b+k}^{(p^2)}(\beta)\in \{0,1\}\,\,\,\, \textrm{for}\,\,\,\, a\in D_k^{(p^2)}.\]
Furthermore, we can prove in the same way as in  Theorem \ref{thm1} that
$s({\beta^a})=1$ for half of the elements in $\mathbb{Z}_{p^2}\backslash \{0\}$ and $s({\beta^a})=0$ for the other half,
that is, there are exactly $\frac{p^2-1}{2}$ elements $a\in \mathbb{Z}_{p^2}\backslash \{0\}$ such that $s({\beta^a})=0$.
Consequently, the size of the set $\{a\in\mathbb{Z}_{p^2}\,:\,s(\beta^a)=0\}$ is $\frac{p^2-1}{2}$,
and thus the linear complexity of sequence (\ref{e1}) is equal to $p^2-\frac{p^2-1}{2}=\frac{p^2+1}{2}$.
\end{proof}

\begin{remark}\label{remark2}
The condition $2^e\equiv 1\,(\bmod\, p^2)$ implies that $2^{p-1}\equiv 1\,(\bmod\, p^2)$ due to $p-1=ef$.
A prime $p$ satisfying $2^{p-1} \equiv 1\,(\bmod\, p^2)$ is called a \textit{Wieferich prime}.
It is shown in \cite{Dorais} that there are only two Wieferich primes, 1093 and 3511, up to $6.7\times10^{15}$.
Therefore, for all non-Wieferich primes $p$ the result in Theorem \ref{thm1} holds. For the known Wieferich prime $p=1093$, one can take $f=2$ or $4$ and thus $e=\frac{p-1}{2}=546$ or $\frac{p-1}{4}=273$.
It is easily checked that $2^e\not\equiv 1\,(\bmod\,p^2)$ and $2^e\not\equiv 1\,(\bmod\,p)$ in both cases, and so the linear complexity
of sequence (\ref{e1}) of period $1093^2$ is equal to its period by Theorem \ref{thm1}.
For another Wieferich prime $p=3511$, $e$ can only take the value $\frac{p-1}{2}=1755$, and one can check that $2^e\equiv 1\,(\bmod\,p^2)$,
so the linear complexity of sequence (\ref{e1}) of period $3511^2$ is equal to $\frac{3511^2+1}{2}$.
\end{remark}

The following two examples illustrate the sequences described in (\ref{e1}) and the main results in Theorem \ref{thm1}.
\begin{example}\label{example2}
	Let $p=5$. Note that $g=2$ is a primitive root modulo $5^2$. If one takes $f=2$ and $e=2$, then the parameter $b$ in the characteristic set given in (\ref{e0})
	can be any integer from $0$ to $9$. By C-language program, for $b=0,1,\,\cdots,\,9$, the corresponding binary sequences of period $5^2$ are as follows:
	\[1110110011000000110011011,\,\,\,\,1010100111100001111001010,\]
	\[1000110111010010111011000,\,\,\,\,1001000111110011111000100,\]
	\[1001011101010010101110100,\,\,\,\,1001001100111111001100100,\]
	\[1101011000011110000110101,\,\,\,\,1111001000101101000100111,\]
	\[1110111000001100000111011,\,\,\,\,1110100010101101010001011.\]
	If one takes $f=4$ and $e=1$, then $b$ can be an integer from $0$ to $19$. For $b=0,1,\,\cdots,\,19$, the corresponding binary sequences of period $5^2$ are as follows:
	\[1111111110101010100000000,\,\,\,\,1011101110101010100010001,\]
	\[1001101110001011100010011,\,\,\,\,1001011110001011100001011,\]
	\[1001011100101010110001011,\,\,\,\,1001001101101010010011011,\]
	\[1001001001001011011011011,\,\,\,\,1001011001011001011001011,\]
	\[1000011001111000011001111,\,\,\,\,1000000001111000011111111,\]
	\[1000000001010101011111111,\,\,\,\,1100010001010101011101110,\]
	\[1110010001110100011101100,\,\,\,\,1110100001110100011110100,\]
	\[1110100011010101001110100,\,\,\,\,1110110010010101101100100,\]
	\[1110110110110100100100100,\,\,\,\,1110100110100110100110100,\]
	\[1111100110000111100110000,\,\,\,\,1111111110000111100000000.\]
	By C-language program, the linear complexities of the above sequences are all equal to 25, which are consistent with the result in Theorem \ref{thm1} since $2\notin D_0^{(p)}$. Furthermore, the experimental results suggest that the above sequences are distinct up to shift equivalence.
\end{example}

\begin{example}\label{example1}
	Let $p=7$. Note that $g=3$ is a primitive root modulo $7^2$. If one takes $f=2$ and $e=3$, then the parameter $b$ in the characteristic set given in (\ref{e0})
	can be any integer from $0$ to $13$. By C-language program, for $b=0,1,\,\cdots,\,13$, fourteen binary sequences of period $7^2$ are as follows:
	\[1101010101000111101000000111111010000111010101010,\]
	\[1001010001000101100101000111010110010111011101011,\]
	\[1000000111000111100100000111110110000111000111111,\]
	\[1000000010000100100101001011010110110111101111111,\]
	\[1000100110000110100100111000110110100111100110111,\]
	\[1000100010001100110111111000000100110011101110111,\]
	\[1010100110011010110110111000100100101001100110101,\]
	\[1010101010111000010111111000000101111000101010101,\]
	\[1110101110111010011010111000101001101000100010100,\]
	\[1111111000111000011011111000001001111000111000000,\]
	\[1111111101111011011010110100101001001000010000000,\]
	\[1111011001111001011011000111001001011000011001000,\]
	\[1111011101110011001000000111111011001100010001000,\]
	\[1101011001100101001001000111011011010110011001010,\]
	and the linear complexities of these sequences are all equal to 46, which are coincident with the result in Corollary \ref{cor1}.
	In addition, one can also verify that the above sequences are distinct up to shift equivalence.
\end{example}

	\begin{remark}\label{remark3} Let $p$ be an odd prime, $f$ be a divisor of $p-1$ and $g$ be a primitive element modulo $p^2$.
		Define
		$
		\widehat{D}_0^{(p^j)} = (g^f) = \{g^{ft} ({\rm\, mod\,} p^j) \,| \, 0\leq t<\varphi(p^j)/f\}
		$ for any integer $j\geq 1$. It is easy to check that $\widehat{D}_0^{(p^j)}$ is a subgroup of $\mathbb{Z}_{p^j}^*$ and it yields the generalized cyclotomic classes of order $f$ with respect to $p^j$ as
		\begin{equation}\label{Eq_Cyclotomy}
		\widehat{D}_i^{(p^j)} = g^i\widehat{D}_0^{(p^j)} = \{g^{i} x ({\rm\, mod\,} p^j) \,| \, x \in \widehat{D}_0^{(p^j)} \}, i=0, 1, \cdots, f-1.
		\end{equation} Then $\mathbb{Z}_{p^2}$ can be partitioned as
		$\mathbb{Z}_{p^2}=\{0\} \cup \bigcup\limits_{i=0}^{f-1}(p\widehat{D}_i^{(p)}\cup \widehat{D}_i^{(p^2)}).$
		Several binary sequences of period $p^2$ were constructed from this generalized cyclotomy \cite{Ding0, Ding4, Du}.
		In the case of $f=2$,
		the generalized cyclotomic binary sequences of period $p^2$ with
		the characteristic set $p\mathbb{Z}_p \cup \widehat{D}_1^{(p^2)}$, and with the characteristic set $\{0\}\cup p\widehat{D}_1^{(p)}\cup \widehat{D}_1^{(p^2)}$
		were introduced in \cite{Ding0, Ding4} and their properties were further investigated in \cite{Yan1}. Recently, the linear complexity of
		the generalized cyclotomic binary sequences with
		the characteristic set $\bigcup\limits_{i=0}^{f/2-1}(p\widehat{D}_i^{(p)}\cup \widehat{D}_i^{(p^2)})$ for $f=6$ was studied in \cite{Du}.
		In this paper, the binary sequences are constructed from a special case of the generalized cyclotomy proposed in \cite{Zeng}, where the subgroup $D_0^{(p^j)}$ (as given in (\ref{Eq_GenCyclotomy}))  is defined in a different manner and
		the cyclotomic classes with repect to $p^j$ have order $d_j=p^{j-1}f$.
		This allows for more flexibility to choose the characteristic sets in (\ref{e0}) of the cyclotomy sequences. As a result, for a fixed primitive element $g$, we derived in total $pf$
		balanced cyclotomic sequences and in the case where $2^{\frac{p-1}{f}}\equiv 1 (\bmod \, p)$ and  $2^{\frac{p-1}{f}}\not\equiv 1 (\bmod \, p^2)$,
the linear complexity of these sequences
is larger than that of those cyclotomic sequences in \cite{Ding0, Yan1, Du}. For cryptographic applications, high linear complexity does not necessarily guarantee the cryptographic strength of a sequence and the linear complexity profile
		should be also taken into consideration. Given a binary sequence $\mathbf{s}_n=(s_0,s_1,\cdots,s_{n-1})$, its linear complexity profile is defined as the sequence $(L_1(\mathbf{s}_n), \cdots, L_n(\mathbf{s}_n))$, where $L_k(\mathbf{s}_n)$ is
		the $k$-th linear complexity of $\mathbf{s}_n$ is  the linear complexity of the sub-sequence
		$(s_0,\cdots, s_{k-1})$  for $1\leq k\leq n$.
		The numerical result indicates that the $k$-th linear complexity of the cyclotomic sequences for $1\leq k\leq p^2-1$ in this paper lies around $\frac{k}{2}$, which is the expected linear complexity of a truly random sequence of length $k$.
		
	\end{remark}

	
	\begin{remark}\label{rem4}
		To analyze the properties of pseudo-random sequences $\textbf{s}_n=(s_0, \cdots, s_{n-1})\in \{0, 1\}^{n}$, Mauduit and S\'{a}rk\"{o}zy in \cite{Mauduit}
		introduced the following measures of pseudorandomness: the \textit{well-distribution} measure of $\textbf{s}_n$ is given by
		$
		W(\textbf{s}_n)=\max\limits_{a, b, t}\big|\sum\limits_{j=0}^{t-1}(-1)^{s_{a+jb}}\big|,
		$ where the maxium is taken over all $a, b, t\in \mathbb{Z}_n$ with $0\leq a \leq a+(t-1)b<n$,
		and the \textit{correlation measure of order $k$} of $\textbf{s}_n$ is given by
		$
		C_k(\textbf{s}_n) = \max\limits_{m, D}\big|\sum\limits_{i=0}^{m-1}(-1)^{s_{i+d_1}+s_{i+d_2}+\cdots+ s_{i+d_k}}\big|,
		$ where the maximum is taken over all $D=(d_1,\cdots, d_k)$ and $m$ with $0\leq d_1<\cdots<d_k\leq n-m$.
		A binary pseudorandom sequence $\textbf{s}_n$ is regarded as ``good" if both $W(\textbf{s}_n)$ and $C_k(\textbf{s}_n)$ (at least for small $k$)
		are ``small" in terms of $n$.
		These two measures were later justified in \cite{Mauduit2} that for almost all truly random sequences $\textbf{s}_n=(s_1,\cdots, s_n)$, both $W(\textbf{s}_n)$ and $C_k(\textbf{s}_n)$ are less than $n^{\frac{1}{2}}\log^c(n)$ for some constant $c$
		as $n\rightarrow \infty$.
		
		It is worth noting that the generalized cyclotomic binary sequences of period $p^2$ in \cite{Ding0} and
		\cite{Ding4, Yan1} satisfy $s_{k}=s_{k+p}$ for each positive integer $k$ with $\textrm{gcd}\,(k,p)=1$ due to $\widehat{D}_i^{(p^2)}(\bmod\,p)=\widehat{D}_i^{(p)}$ for $i\in \{0,1\}$.
		This leads to a large  correlation measure of order $2$.
		In contrast, the cyclotomic sequences constructed from the generalized cyclotomy in (\ref{Eq_GenCyclotomy}) do not have such a property.	
		We calculate the well-distribution measure and correlation measure of the cyclotomic sequences of period $p^2$ constructed in this paper.
		The numerical results show that in the case $f=2$, the correlation measure of order $2$ of all  sequences in this paper are smaller than that of the sequences in \cite{Ding0, Ding4, Yan1}. Moreover,
		both the well-distribution measure and the correlation measure of order $k$ for $1\leq k \leq 4$ of all sequences are  below the upper bound $p\log(p^2)$, which indicates that the cyclotomic sequences
		in this paper are potentially  `` good " in terms of these pseudorandomness measures.
	\end{remark}

We finally remark that it is natural to  further consider the similar construction of new generalized cyclotomic sequences of period $p^m$ for an odd prime $p$ and an integer $m\geq 3$. Let
\[C_0=\bigcup\limits_{j=1}^{m} \bigcup\limits_{i=d_j/2}^{d_j-1} p^{m-j}D_{(i+b)\,(\textrm{mod}\,d_j)}^{(p^j)},\,\,\,\,
C_1=\bigcup\limits_{j=1}^{m} \bigcup\limits_{i=0}^{d_j/2-1} p^{m-j}D_{(i+b)\,(\textrm{mod}\,d_j)}^{(p^j)}\cup \{0\} ,\]
one can define the generalized cyclotomic binary sequences of period $p^m$ as
\begin{equation}\label{e8}
s_i = \left\{ {\begin{array}{*{20}c}
	{0}, \hfill & {i\,(\textrm{mod}\,p^m) \in C_0 ,} \hfill  \\
	{1}, \hfill & {i\,(\textrm{mod}\,p^m) \in C_1,} \hfill  \\
	\end{array}} \right.
\end{equation}
We observe, on basis of numerical results, that the main result in Theorem 1 also holds for the sequences defined by (\ref{e8}).
However, we fail to prove it using the same method as in this paper.
In fact, we can extend the results in Proposition \ref{lem6}, Lemmas \ref{lem1}, \ref{lem7} and the sufficiency in Lemma \ref{lem8} to the generalized cyclotomic binary sequences of period $p^m$.
But since the order of the generalized cyclotomic classes with respect to $p^j$ depends on the value of $j$, we can't deduce a contradiction
as in the proof of  the necessity in Lemma
\ref{lem8} by establishing a system of equations and discussing the number of its solutions. We provide the result as a conjecture below and cordially invite interested readers to attack this problem.

\begin{conj}
	Let $p$ be a non-Wieferich odd prime and let $\mathbf{s}^\infty$ be a generalized cyclotomic binary sequence of period $N=p^m$ defined by (\ref{e8}).
	Then its linear complexity is given by
	\[L(\mathbf{s}^{\infty}) = \left\{ {\begin{array}{*{20}c}
		p^m-\frac{{p-1}}{2}-\delta(\frac{p^m+1}{2}), \hfill & {2 \in D_0^{(p)}}, \hfill  \\
		p^m-\delta(\frac{p^m+1}{2}), \hfill & {2 \notin D_0^{(p)}}, \hfill  \\
		\end{array}} \right.\]
	where $\delta(t)=1$ if $t$ is even and $\delta(t)=0$ if $t$ is odd.
\end{conj}

\section{Conclusion}
In this paper, new classes of almost balanced binary sequences of period $p^2$ were constructed
via generalized cyclotomic classes proposed in \cite{Zeng}. The linear complexity of these sequences is also determined.
The results show that the proposed sequences can have large linear complexity $p^2-\frac{p-1}{2}$ or $p^2$ if $2^e\not\equiv 1\,(\bmod\,p^2)$,
which is very close to the period. In fact, for $p<6.7\times10^{15}$ only one pair $(p,e)$ satisfying $2^e\equiv 1\,(\bmod\,p^2)$.
Since the parameters $e$ and $f$ are not uniquely determined, and the parameter $b$ in the characteristic set could
be any integers in the range of $0$ to $pf-1$, our construction can generate a great number of binary sequences with large linear complexity.
In addition, the numerical result (for small odd primes $p$) indicates that
the cyclotomic sequences constructed in this paper are  ``good"  in terms of
the linear complexity profile, the well-distribution measure and the correlation measure of pseudorandom sequences.
To estimate these measures of the cyclotomic sequences theoretically is an interesting problem and we will work on it in future research.

\end{document}


\title{Corrigendum to ``New generalized cyclotomic binary sequences of period $p^2$"}
\author{Zibi Xiao
\thanks{Z. Xiao and X. Zeng are with the Faculty of Mathematics and Statistics, Hubei Key Laboratory of Applied Mathematics,
Hubei University, Wuhan 430062, Hubei, China. Email: holly\_xzb@126.com, xzeng@hubu.edu.cn},
Xiangyong Zeng, Chunlei Li
\thanks{C. Li and T. Helleseth are with the Department of Informatics, University of Bergen,
Bergen {\rm N-5020}, Norway. Email: chunlei.li@uib.no, tor.helleseth@uib.no},
and
Tor Helleseth
}

\date{}
\maketitle
\begin{quote}
{\small {\bf Abstract:}} We pointed out that the proof of Lemma 6 in [1] only holds for $f=2$ and gave a proof
for any $f=2^r$ when $p$ is a non-Wieferich prime.


\end{quote}
\section*{}
The following sentence in the proof of Lemma 6 of the reference \cite{XZ} is not accurate:
\begin{quotation}
	``Note that each polynomial $U_n(x)$ with $0\leq n\leq p-1$ satisfies $U_n(\beta)=0$ for an arbitrary primitive $p^2$-th root of unity $\beta$. This implies the degree of $U_n(x)$ is no less than the number of primitive $p^2$-th roots of unity over $\mathbb{F}_2$."
\end{quotation}

We remark that the above statement holds only for the case $f=2$ when $p$ is an arbitrary odd prime.
As a consequence, Theorems 1 and 2  hold only for the case $f=2$.

\smallskip

We also remark that when $p$ is a non-Wieferich prime with $2^{p-1}\not\equiv 1 \pmod{p^2}$, the statements of Lemma 6 and Theorem 1 indeed hold for any $f=2^r$ with $r\geq 1$, which is also remarked in \cite{VL}.
In this case, the last two paragraphs of the proof of Lemma 6 in \cite{XZ} shall be revised as follows:

\begin{quotation}
	``Observe that for $n=0, 1,,\cdots, p-1$,
	$$U_n\hspace{-10pt}\pmod{p} = \bigcup\limits_{i=0}^{\frac{f}{2}-1}\big(D_{i+h}^{(p)}\cup D_{i+h+\frac{f}{2}}^{(p)}\big)=\mathbb{Z}_{p}^*.$$
	Thus, by letting $\bar{t}=t \pmod{p}$, we have
	$$
	0=U_n(\beta) = \sum\limits_{t\in U_n}\beta^t=\sum\limits_{t\in U_n}(\beta^p)^{\frac{t-\bar{t}}{p} }\beta^{\bar{t}} =\sum\limits_{t\in U_n}(\beta_1)^{\frac{t-\bar{t}}{p} }\beta^{\bar{t}} = \sum\limits_{i=1}^{p-1}c_i\beta^{i},
	$$
	where $\beta_1=\beta^p$ is a $p$-th primitive root of unity over $\mathbb{F}_2$ and $c_i\in \mathbb{F}_2(\beta_1)$. This means that the dimension of the extension field $\mathbb{F}_2(\beta)$ over $\mathbb{F}_2(\beta_1)$,
	denoted by $[\mathbb{F}_2(\beta): \mathbb{F}_2(\beta_1)]$ as usual, is at most $p-1$.
	On the other hand,
	for a non-Wieferich prime $p$ with $2^{p-1}\not\equiv 1 \hspace{-2pt}\pmod{p^2}$, it can be easily shown that
	$[\mathbb{F}_2(\beta): \mathbb{F}_2]={\rm ord}_p(2)p$ and $[\mathbb{F}_2(\beta_1): \mathbb{F}_2]={\rm ord}_p(2)$, where ${\rm ord}_p(2)$ is the order of $2$ modulo $p$.
	This implies $[\mathbb{F}_2(\beta): \mathbb{F}_2(\beta_1)]=p$, which is a contradiction.

\end{quotation}

%